\documentclass[twocolumn,floatfix,showpacs,aps,prd]{revtex4}

\usepackage{graphicx}
\usepackage{dcolumn}
\usepackage{amsmath}
\usepackage{units}

\input{babarsym}

\def\mpp {\ensuremath{m_{\phi\phi}}\xspace}
\def\btoppk  {\ensuremath{B \to \phi \phi K}\xspace}
\def\bptoppk  {\ensuremath{B^+ \to \phi \phi K^+}\xspace}
\def\bztoppk  {\ensuremath{B^0 \to \phi \phi K^0}\xspace}

\def\btoeck  {\ensuremath{B \to \eta_c K}\xspace}

\def\ectopp   {\ensuremath{\eta_c \to \phi \phi}\xspace}

     \def\ppkpm {\ensuremath{\phi \phi K^\pm}\xspace}
     \def\pkkkpm {\ensuremath{\phi KK K^\pm}\xspace}
     \def\kkkkkpm {\ensuremath{KKKK K^\pm}\xspace}
     \def\fzpkpm {\ensuremath{f_0 \phi K^\pm}\xspace}
     \def\fzkkkpm {\ensuremath{f_0 KK K^\pm}\xspace}

     \def\ppks {\ensuremath{\phi \phi K^0_s}\xspace}
     \def\pkkks {\ensuremath{\phi KK K^0_s}\xspace}
     \def\kkkkks {\ensuremath{KKKK K^0_s}\xspace}
     \def\fzpks {\ensuremath{f_0 \phi K^0_s}\xspace}
     \def\fzkkks {\ensuremath{f_0 KK K^0_s}\xspace}

     \def\btoeck  {\ensuremath{B \to \eta_c K}\xspace}

     \def\ectopp   {\ensuremath{\eta_c \to \phi \phi}\xspace}

\newcommand{\BABARPubYear}    {07}
\newcommand{\BABARPubNumber}  {x}
\newcommand{\SLACPubNumber} {x}

\def\figurebox#1#2#3{%
    \def\arg{#3}%
    \ifx\arg\empty
    {\hfill\vbox{\hsize#2\hrule\hbox to #2{\vrule\hfill\vbox to #1{\hsize#2\vfill}\vrule}\hrule}\hfill}%
    \else
    {\hfill\epsfbox{#3}\hfill}%
    \fi}

\long\def\inst#1{\par\nobreak\kern 4pt\nobreak
    {\it #1}\par\vskip 10pt plus 3pt minus 3pt}

\begin{document}

\preprint{\babar-PUB-\BABARPubYear/\BABARPubNumber}
\preprint{SLAC-PUB-\SLACPubNumber}

\begin{flushleft}
 BABAR-PUB-11/003;  SLAC-PUB-14473 \\
\end{flushleft}

\title{
  { \Large \bf \boldmath Measurements of branching fractions and $CP$ asymmetries
            and studies of angular distributions for
             $B \to \phi \phi K$ decays}
}
%
\author{J.~P.~Lees}
\author{V.~Poireau}
\author{E.~Prencipe}
\author{V.~Tisserand}
\affiliation{Laboratoire d'Annecy-le-Vieux de Physique des Particules (LAPP), Universit\'e de Savoie, CNRS/IN2P3,  F-74941 Annecy-Le-Vieux, France}
\author{J.~Garra~Tico}
\author{E.~Grauges}
\affiliation{Universitat de Barcelona, Facultat de Fisica, Departament ECM, E-08028 Barcelona, Spain }
\author{M.~Martinelli$^{ab}$}
\author{D.~A.~Milanes$^{a}$}
\author{A.~Palano$^{ab}$ }
\author{M.~Pappagallo$^{ab}$ }
\affiliation{INFN Sezione di Bari$^{a}$; Dipartimento di Fisica, Universit\`a di Bari$^{b}$, I-70126 Bari, Italy }
% ...
\collaboration{The \babar\ Collaboration}
\noaffiliation

 \collaboration{The \babar\ Collaboration}

\date{\today}

\begin{abstract}
   We present branching fraction and $CP$ asymmetry measurements
   as well as angular studies of
   $B \to \phi \phi K$ decays
   using $464\times 10^6$ \BB\ events collected by the \babar\ experiment.
   The branching fractions are measured in the
   $\phi \phi$ invariant mass range below the $\eta_c$ resonance ($m_{\phi\phi}<2.85$ GeV).
   We find
     ${\cal B}(B^+ \to \phi \phi K^+) = (5.6 \pm 0.5 \pm 0.3)\times 10^{-6}$
   and
     ${\cal B}(B^0 \to \phi \phi K^0) = (4.5 \pm 0.8 \pm 0.3)\times 10^{-6}$,
   where the first uncertaintiy is statistical and the second systematic.
   The measured direct $CP$ asymmetries for the $B^\pm$ decays are
     $A_{CP} = -0.10 \pm 0.08 \pm 0.02$ below the $\eta_c$ threshold ($m_{\phi\phi}<2.85$ GeV)
   and
     $A_{CP} = 0.09 \pm 0.10 \pm 0.02$ in the $\eta_c$ resonance region
     ($m_{\phi\phi}$ in [2.94,3.02] GeV).
   Angular distributions are consistent with $J^P = 0^-$ in the $\eta_c$ resonance region
   and favor $J^P = 0^+$ below the $\eta_c$ resonance.
\end{abstract}


\maketitle


  The violation of $CP$ symmetry is a well-known requirement for the matter-antimatter
  imbalance of the universe~\cite{sakarov}.
  The \babar~\cite{babarnim} and Belle~\cite{bellenim} experiments at the high-luminosity $B$ factories,
  PEP-II~\cite{pepii} and KEKB~\cite{kekb}, have made numerous
  $CP$ asymmetry measurements using datasets
  two orders of magnitude larger than their predecessors.
  All of these measurements are consistent with a single source of $CP$ violation --
  the complex phase within the CKM quark mixing matrix of the Standard Model~\cite{km}.
  However, with the small amount of $CP$ violation from the CKM matrix, it
  is difficult to explain the matter-antimatter asymmetry of the
  universe~\cite{matter-antimatter-asym}.
  This motivates searches for new sources of $CP$ violation.

  A method to search for new sources of $CP$-violating phases is to measure
  $CP$ asymmetries in hadron decays that are forbidden at the tree level~\cite{grossman}.
  Since the leading decay amplitude is a one-loop process, contributions
  within the loop from virtual non-Standard-Model particles cannot be excluded.
  The quark interactions with the non-Standard-Model particles can introduce
  new $CP$ violating phases in the decay amplitude, which can lead to
  observable non-zero $CP$ asymmetries.
  Decays of $B$ mesons with a $b\to s \bar s s$ transition have been extensively
  studied for this reason.

  The three body \btoppk decay is a one-loop ``penguin''
  $b\to s \bar s s$ transition.
  This final state can also occur through the tree-level decay
  \btoeck ,
  followed by
  \ectopp ,
  where
  the $B$ decay is a $b\to c\bar c s$ transition.
  If the $\phi\phi$ invariant mass \mpp in the
  three-body \btoppk  decay is close to the
  $\eta_c$ resonance, the tree and penguin amplitudes may interfere.
  Within the Standard Model, the relative weak phase between these
  amplitudes is $\arg(V_{tb}V_{ts}^*/V_{cb}V_{cs}^*) \approx 0$, so
  no $CP$ violation is expected from the interference.
  However, new physics contributions to the penguin loop in the
  \btoppk
  decay could introduce a non-zero relative $CP$
  violating phase, which may then produce a significant
  direct $CP$ asymmetry~\cite{hazumi}.
  Measurement of a significant, non-zero direct $CP$ asymmetry
  would be an unambiguous
  sign of new physics.
  A previous measurement of the direct $CP$ asymmetry~\cite{belle-ppk}
  was consistent with zero, but was also limited by a large statistical
  uncertainty.
  The $B^+$ and $B^0$ branching
  fractions have been previously measured~\cite{belle-ppk}\cite{old-babar-ppk}
  to be a few times $10^{-6}$.
  Theoretical predictions of the branching fractions are of the
  same order \cite{fajfer}\cite{chen}.


  \section{Dataset and Detector description}

  We present measurements of the \bptoppk and \bztoppk
  branching fractions~\cite{chargeconj} and direct $CP$ asymmetry
  $A_{CP} \equiv [N(B^-)-N(B^+)]/[N(B^-)+N(B^+)]$
  as well as studies of angular distributions
  performed using $464 \times 10^6$
  \BB
  pairs collected by the \babar\ experiment
  at the SLAC National Accelerator Laboratory.
  The direct $CP$ asymmetry is measured both below and within the
  $\eta_c$ resonance region of the $\phi\phi$ invariant mass
  with these regions defined as $m_{\phi\phi}<2.85$~GeV and $m_{\phi\phi}$ within
  [2.94, 3.02]~GeV, respectively~\cite{c=1}.
  The branching fractions are measured in the \mpp region below the
  $\eta_c$ resonance ($m_{\phi\phi}<2.85$~GeV).

  The \babar\ detector is described in detail elsewhere~\cite{babarnim}.
  What follows is a brief overview of the main features of the detector.
  The detector has a roughly cylindrical geometry, with the axis along
  the beam direction.
  The trajectories, momenta, and production vertices of charged particles
  are reconstructed from position measurements made by a silicon
  vertex tracker (SVT) and a 40-layer drift chamber (DCH).
  The SVT consists of 5 layers of double-sided silicon strip detectors
  which provide precision position measurements close to the beam interaction region.
  Both the SVT and DCH measure the specific energy loss ($dE/dx$) along
  the charged particle trajectory, which is used to infer the particle
  mass from the velocity dependence of the energy loss and the momentum
  measurement.
  The tracking system is inside a uniform 1.5~T magnetic field provided
  by a superconducting solenoid.
  Outside the tracking system, an array of quartz bars coupled with an array of phototubes
  (DIRC) detects the Cherenkov light produced when a charged particle travels
  through the quartz bars.
  The measured Cherenkov angle is used to infer the particle mass from the velocity
  dependence of the Cherenkov angle and the measured momentum.
  The energies of photons and electrons are determined from the measured light produced
  in electromagnetic showers inside a CsI crystal calorimeter (EMC).
  Gaps in the iron of the magnet flux return are instrumented with resistive plate chambers
  and limited streamer tubes, which are used to identify muons and neutral hadrons (IFR).

  We use Monte Carlo (MC) samples to determine the signal selection efficiency.
  The MC events are generated with EvtGen~\cite{evtgen} and simulated using
  \textsc{Geant4}~\cite{geant4}.


  \section{Event selection}



     We select
     events containing multiple hadrons by requiring at least three charged tracks
     in the event and
     the ratio of the second to zeroth Fox-Wolfram\cite{fox-wolfram} moments $R_2$ to be less than 0.98.

   %
   %
     Charged kaon candidates are required to pass a selection based on a likelihood ratio
     which uses the SVT and DCH $dE/dx$ and the DIRC Cherenkov angle measurements
     as inputs to the likelihood.
     The ratio is defined as $R_{hh'} \equiv {\cal L}_h / ( {\cal L}_h + {\cal L}_{h'})$, where
     $h$ and $h'$ are $K$, $\pi$, or $p$.
     The minimum kaon selection criterion is $R_{K\pi}>0.2$ or $R_{p\pi}>0.2$.
     This selection has an efficiency greater than 98\% for kaons
     and a pion efficiency of less than 15\%
     below a lab momentum of 2.5~GeV.
     Candidate $\phi \to K^+K^-$ decays are constructed from oppositely-charged kaon
     candidates with an invariant mass in the range of 0.987 to 1.2~GeV.
     At least one of the kaons in each $\phi \to K^+K^-$ candidate must also satisfy the
     more stringent criteria of $R_{K\pi}>0.8176$ and $R_{Kp}>0.018$,
     which has an efficiency greater than 90\% for kaons and a pion efficiency of less than 3\%
     below a lab momentum of 2.5~GeV.


     Candidate $K^0_S \to \pi^+\pi^-$ decays are constructed from oppositely-charged
     pion candidates with an invariant mass in the range of 0.486 to 0.510~GeV.
     The pion tracks are fit to a common vertex.
     The $\chi^2$ probability of the vertex fit must be greater than 0.001.
     The typical experimental resolution on the measured $K^0_S$ flight length in the
     plane transverse to the beam is around 0.2~mm or less.
     We require the transverse flight length to be at least 2~mm.


     Candidate \btoppk\ decays are constructed from
     pairs of $\phi$ candidates that do not share any daughters
     and either a $K^\pm$ or a $K^0_S$ candidate.
     The $\phi$ and $K^\pm$ candidates are constrained to a common vertex.
     We reject
     combinatoric background by requiring the $B$ candidate to have
     kinematics consistent with $\Upsilon(4S) \to \BB$ using two standard
     variables: $m_{\rm ES}$ and $\Delta E$.
     The energy-substituted $B$ mass is defined as $m_{\rm ES} \equiv \sqrt{ E_{\rm beam}^{*2} - p^{*2} }$,
     where $E^*_{\rm beam}$ and $p^{*}$ are the beam energy ($\sqrt{s}/2$) and the reconstructed $B$
     momentum, both in the center-of-mass (CM) reference frame.
     The energy difference is defined as $\Delta E \equiv E^* - E_{\rm beam}^{*}$,
     where $E^*$ is the reconstructed $B$ energy in the CM frame.
     We require $m_{\rm ES}$ and $\Delta E$ to be within [5.20, 5.29]~GeV and 
     [$-0.1, 0.1$]~GeV, respectively.
     The experimental resolution is about 2.7 MeV for $m_{\rm ES}$
     and 15 MeV for $\Delta E$.

     The $m_{\rm ES}$ interval includes a large ``sideband'' region
     below the area where the signal is concentrated near the $B$ mass.
     The $\Delta E$ interval also is wide enough to include sideband regions
     where the signal probability is very low.
     Including events in the sideband regions
     enables us to determine the probability density functions (PDFs) of the
     combinatoric background directly in the maximum likelihood (ML)
     fits of the data.

     About 7\% of events in signal Monte Carlo samples have more than
     one \btoppk\ candidate.
     If there are multiple $B^+\to \phi\phi K^+$ candidates in a single event,
     we select the $B^+$ candidate with the smallest mass $\chi_m^2$ defined as
     $\sum_i \frac{(m_i - m_{0})^2}{\sigma^2}$, where
     the sum $i$ is over the two $\phi$ candidates,
     $m_i$ ($m_{0}$) is the reconstructed (nominal) $\phi$ mass,
     and $\sigma$ is the RMS of the reconstructed $m_{\phi}$ distribution
     for properly reconstructed $\phi$ candidates.
     If there are more than one $B^+$ candidates that use the same two
     $\phi$ candidates, we choose the $B^+$ candidate with the highest
     quality $K^+$ identification for the $K^+$ from the $B^+$ decay.
     If the quality level of the $K^+$ identification is the same for
     these $B^+$ candidates, we choose the $B^+$ candidate with the
     highest vertex $\chi^2$ probability.
     For events with multiple $B^0\to \phi\phi K^0_S$, the sum for
     $\chi_m^2$ includes the $K^0_S \to \pi^+\pi^-$, and we choose
     the $B^0$ with the smallest $\chi_m^2$.
     For both the $B^+$ and $B^0$ decay modes, 
     the probability that the the algorithms described
     above choose the correct candidate is about 87\%.


     The reconstruction and selection efficiencies for events with
     $m_{\phi\phi} < 2.85$ GeV are determined from Monte Carlo samples to be
     28.0\% and 22.5\% for the $B^+$ and $B^0$ modes, respectively.


     We use control samples of $B \to D_s D$ decays where $D_s \to \phi \pi$,
     $D\to K\pi$, and $\phi \to K^+K^-$ to determine corrections to the
     $B\to \phi\phi K$ signal probability density function parameters
     determined from Monte Carlo samples in the maximum likelihood fits described below.

     \subsection{ Continuum Background }

\begin{figure}
     \begin{center}
         \includegraphics[width=\linewidth]{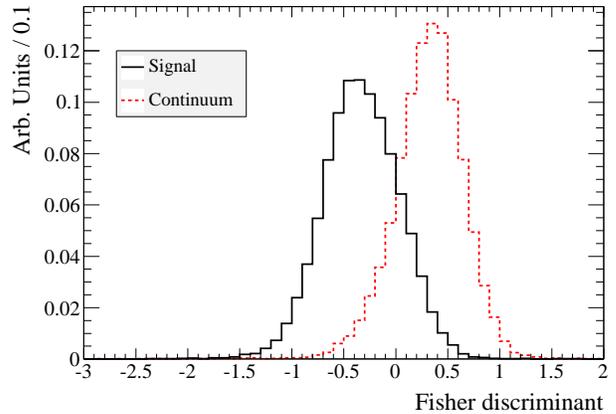}
       \caption{ Distributions of the Fisher discriminant ${\cal F}$
               for $B\to \phi \phi K$ signal (solid black) and
               $e^+e^- \to q\bar q$ continuum (dashed red) Monte
               Carlo simulation.
                 }
       \label{fig:fisher}
     \end{center}
   \end{figure}

     The events that pass the selection above with at least one $B$ candidate are
     primarily background events from the continuum ($e^+e^- \to q\bar q$ with
     $q = u,d,s,c$).
     We reduce this background by using a Fisher discriminant (${\cal F}$),
     which is the linear combination of seven variables and is optimized for maximum
     separation power of signal and the continuum background.
     The seven variables are listed below.
     These variables are commonly used by the \babar\ experiment in analyses of
     charmless $B$ decays, where the primary background is from continuum
     events.
     They take advantage of aspects of the production distributions and event
     topologies of \BB versus continuum $q\bar q$ production events. \\
     %
     %
     \begin{itemize}
       \item {\boldmath $|\Delta t/\sigma_{\Delta t}|$:} the absolute value of the
         reconstructed proper time difference between the two $B$ decays divided by
         its uncertainty~\cite{dtandft}.
       \item {\boldmath $|FT|$:} the absolute value of the standard \babar\ flavor
         tagging neural network output~\cite{dtandft}.
       \item {\boldmath $|\cos\theta^*_{\rm th}|$:} the absolute value of the cosine
         of the angle between the $B$ candidate thrust axis and the thrust axis of the
         rest of the event computed in the CM frame.
         The thrust axis is the direction that maximizes the scalar sum of the projection
         of the track momenta on that direction.
       \item {\boldmath $|\cos\theta^*_{B{\rm thr}}|$:} the absolute value of the cosine
         between the thrust axis of the $B$ candidate and the beam axis in the
         CM frame.  Signal events have a uniform distribution in this variable, while
         continuum background follows a $|1+\cos^2\theta|$ distribution, where
         $\theta$ is the angle between the thrust direction and the beam axis.
       \item {\boldmath $|\cos\theta^*_B|$:} the absolute value of the cosine 
         of the angle between the $B$ direction and the beam axis in
         the CM frame.
         The angular distribution of the signal follows a $\sin^2 \theta^*_B$ distribution,
         while the continuum background is uniformly distributed.
       \item {\boldmath $L_0$ and $L_2$}: The zeroth and second angular moments of
         the momentum flow of the rest of the event about the $B$ thrust axis, defined as
         $L_j \equiv \sum_i p_i |\cos \theta_i|^j$, where the angle
         $\theta_i$ is the angle between track $i$ and the $B$ thrust axis
         and the sum excludes the daughters of the $B$ candidate.
         The calculations are done in the CM frame.
     \end{itemize}
  Distributions of ${\cal F}$ for signal and continuum MC samples
  are shown in Fig.~\ref{fig:fisher}.
  The Fisher discriminant ${\cal F}$ is used as one of several variables
  in the maximum likelihood fits described below.

  \subsection{Peaking Backgrounds}


  The ultimate detected state of our $B\to \phi\phi K$ signal decay is five kaons.
  In addition to the $\phi$ resonance, there may be contributions to
  each $K^+K^-$ pair either from other intermediate $K^+K^-$ resonances,
  such as the $f_0(980)$, or from non-resonant $K^+K^-$ contributions.
  We use the $K^+K^-$ mass sidebands for each $\phi$ candidate to determine
  the amount of $B$ mesons that decay to the detected five-kaon state
  (which we denote $B\to 5K$)
  that are not coming from $B\to \phi\phi K$.
  The specific $B$ decays that we consider as backgrounds are
  $B\to \phi K^+K^-K$,
  $B\to K^+K^-K^+K^-K$,
  $B\to f_0 \phi K$, and
  $B\to f_0 K^+K^-K$.
  The branching fractions for these decays are currently unknown.
  We call these $B$ decays ``peaking backgrounds'' because properly reconstructed
  $B$ candidates are indistinguishable from our $B\to \phi\phi K$ signal in
  the $m_{\rm ES}$, $\Delta E$, and ${\cal F}$ variables.


  We perform unbinned extended maximum likelihood fits to determine the signal
  and combinatoric background yields and, in some cases, the charge asymmetry.
  All of the fits use the product of one-dimensional
  PDFs of $m_{\rm ES}$, $\Delta E$, and ${\cal F}$ in the likelihood.
  For the $B\to \phi\phi K$ branching fraction measurements, we also include
  PDFs for the invariant mass of each $\phi \to K^+K^-$ candidate ($m_{\phi 1}$
  and $m_{\phi 2}$).



  As a first step, we divide the $m_{\phi 1}$ vs. $m_{\phi 2}$ plane~\cite{phi-assignment}
  in the range of 0.987 to 1.200 GeV
  into five mutually exclusive zones.
  We fit for the $B \to 5K$ yield in each zone using only
  $m_{\rm ES}$, $\Delta E$, and ${\cal F}$ in the likelihood.
  The zones are based on various combinations of the $\phi$ signal and sideband regions,
  which are defined as: {\tt Low-SB} [0.987,1.000] GeV, {\tt phi-signal} [1.00,1.04] GeV,
  and {\tt High-SB} [1.04,1.20] GeV.
  Each of the five zones is chosen so that either the $B\to \phi\phi K$ signal or
  one of the four peaking $B$ backgrounds is concentrated in the region.
  We compute the number of peaking background events within the $m_{\phi}$
  range used for the branching fraction fit by using the results of the
  five zone fits as described below.

\begin{figure}
     \begin{center}
         \includegraphics[width=\linewidth]{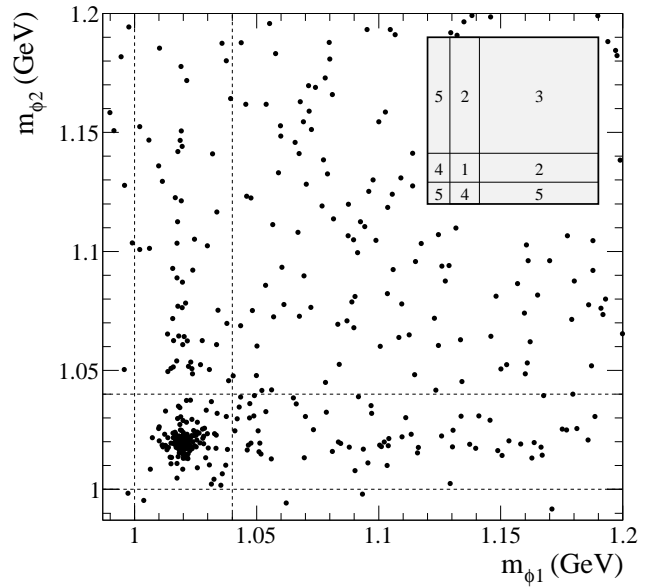}
       \caption{ Data distribution of $B^+ \to 5K$ events in the $m_{\phi 2}$ vs. $m_{\phi 1}$ plane
                 for events with $m_{\phi\phi} < 2.85$ GeV.
                 To enhance the $B\to 5K$ signal for the figure, we have required
                 $m_{\rm ES}>5.27$ GeV, $|\Delta E|<0.040$ GeV,
                 and 
                 ${\cal F}<0.0$.  
                 The efficiency of these additional requirements, relative
                 to the nominal selection, is about 70\% for the signal.
                 The inset shows the definition of the five zones.
                 }
       \label{fig:mphi1vsmphi2}
     \end{center}
   \end{figure}

  Figure~\ref{fig:mphi1vsmphi2} shows the distribution of events in
  the $m_{\phi 2}$ vs. $m_{\phi 1}$ plane for the selected
  $B^+\to 5K$ candidates in the data.
  To enhance the $B^+\to 5K$ signal for the figure,
  we have required $m_{\rm ES}>5.27$ GeV,
  $|\Delta E|<0.040$ GeV, and
  ${\cal F}<0.0$.
  The inset of the figure shows the definition of the five zones.
  A concentration of events in the {\tt phi-signal} region for both $\phi$
  candidates (zone 1) is clearly evident.
  The region defined as {\tt phi-signal} combined with {\tt High-SB} for either
  $\phi$ candidate (zone 2) contains the largest fraction of the
  $B\to \phi K^+K^-K$ mode, although the $B\to \phi\phi K$ signal also
  populates this region due to cases where one $\phi$ is mis-reconstructed.
  The zone where the invariant mass of both $\phi$ candidates is in
  the {\tt High-SB} region (zone 3) contains the largest concentration
  of the non-resonant $B\to K^+K^-K^+K^-K$ mode.
  Zones 4 and 5 contain a large fraction of the $B\to f_0 \phi K$ and
  $B\to f_0 K^+K^- K$ modes, respectively, and very small fractions of
  the other three modes.

  Monte Carlo samples for the five $B$ decay modes (signal plus four
  peaking background modes) are used to determine the fraction of
  events in each zone ($i$) for each decay mode ($j$), which we
  denote with the matrix $f_{ij}$.
  The total $B\to 5K$ yield ($n_i$) is determined for each
  zone $i$ using five separate maximum likelihood fits of the data.
  The yield for each $B$ decay mode ($N_j$) and the amount of each
  mode $j$ in zone $i$  ($n_{ij}$) can be determined from
  \begin{equation}
    N_j = \sum_i f^{-1}_{ij} n_i \ \ \ \ {\rm and} \ \ \ n_{ij} = f_{ij} N_j .
  \end{equation}
  Zone 1 corresponds to the $m_{\phi}$ range used in the branching fraction
  maximum likelihood fit.

  \subsection{ Maximum Likelihood Fits }

  The extended maximum likelihood fits in the five zones determine the
  $B\to 5K$ signal and combinatoric background yields in each zone.
  The $B\to 5K$ signal is split into properly reconstructed and
  misreconstructed (``self-crossfeed'') components, with the
  self-crossfeed fraction fixed.
  The self-crossfeed component is defined as events where
  a true $B\to 5K$ decay is present in the event, but one
  or more tracks used in the reconstructed $B$ are either
  from the other $B$ in the event or not real.
  In zone 1,
  the self-crossfeed fraction for $B\to \phi\phi K$ decays is
  around 7\%.
  %


  The properly reconstructed $B\to 5K$ signal component is described by
  the following PDFs:
  a Crystal Ball function~\cite{CB-function} for $m_{\rm ES}$,
  the sum of three Gaussians for $\Delta E$,
  and the sum of a bifurcated Gaussian and a Gaussian for ${\cal F}$.
  The Crystal Ball function is a Gaussian modified to have an extended
  power-law tail on the low side.
  The $B\to 5K$ signal PDF parameters are determined from MC
  samples with corrections to the $m_{\rm ES}$ and $\Delta E$ core
  mean and width parameters from the $B\to D_s D$ control samples.
  The mean corrections are $0.04 \pm 0.11$ MeV and $-3.5 \pm 0.8$ MeV
  for $m_{\rm ES}$ and $\Delta E$, respectively.
  The width scale factors are $1.10 \pm 0.04$ and $1.04 \pm 0.05$
  for $m_{\rm ES}$ and $\Delta E$, respectively.
  The combinatoric background is described by the following PDFs:
  an empirical threshold function~\cite{Argus-function} for $m_{\rm ES}$,
  a first-order polynomial for $\Delta E$,
  and the sum of two Gaussians for ${\cal F}$.
  Most of the combinatoric background PDF shape parameters are determined
  in the fits.

  The results of the five zone fits for the $B^+$ and $B^0$ modes are
  given in Tables~\ref{tab:b+zone-fits} and~\ref{tab:b0zone-fits}, respectively,
  in the appendix.
  The $B\to \phi\phi K$ signal is observed in both the $B^+$ and $B^0$ samples.
  The $B^0 \to \phi\phi K^0$ decay has not been observed previously.
  The $B$ yield in zone 2 for the $B^+$ mode is significant, but about half of
  this is due to misreconstructed $B\to \phi\phi K$ signal.
  The computed $B^+\to \phi K^+K^-K^+$ and $B^+\to K^+K^-K^+K^-K^+$ yields
  are positive, but the significance is less than two standard deviations.
  There is no evidence of either $B\to f_0\phi K$ or $B \to f_0 K^+K^-K$.
  The branching fraction maximum likelihood fits use the $m_{\phi}$ range
  that corresponds to zone 1.
  We fix the yield of each of the four peaking background modes to the
  zone 1 value in Table~\ref{tab:b+zone-fits} or~\ref{tab:b0zone-fits}
  for the branching fraction fit described below.


   \section{Branching Fraction Analysis }

   The maximum likelihood fit used to measure the $B\to \phi \phi K$ yield
   below the $\eta_c$ resonance for the branching fraction
   measurement restricts
   the event selection with $m_{\phi\phi} < 2.85$ GeV and
   $m_{\phi}$ within [1.00,1.04] GeV, which corresponds to zone 1 in
   the peaking background discussion above.
   The fit components are $B\to \phi \phi K$ signal,
   combinatoric background,
   and the four peaking backgrounds.

   In addition to $m_{\rm ES}$, $\Delta E$, and ${\cal F}$, PDFs for
   $m_{\phi 1}$ and $m_{\phi 2}$ are included in the likelihood function.
   For each fit component, each $\phi$ candidate has a PDF that is the sum
   of a properly reconstructed $\phi \to K^+K^-$ decay,
   given by a relativistic Breit-Wigner function,
   and a misreconstructed $\phi$, described by a first-order polynomial.
   The $m_{\phi 1}$ and $m_{\phi 2}$ PDFs are combined in a way that
   is symmetric under $1 \leftrightarrow 2$ exchange and
   takes into account the fractions of events where both $\phi$ candidates are
   properly reconstructed, one $\phi$ is misreconstructed, and
   both $\phi$ candidates are misreconstructed.

   In addition to the signal and combinatoric background yields,
   the charge asymmetry for the signal and combinatoric background
   components and most of the combinatoric background PDF parameters
   are determined in the fit.

\begin{figure}[h!]
     \begin{center}
         \includegraphics[width=\linewidth]{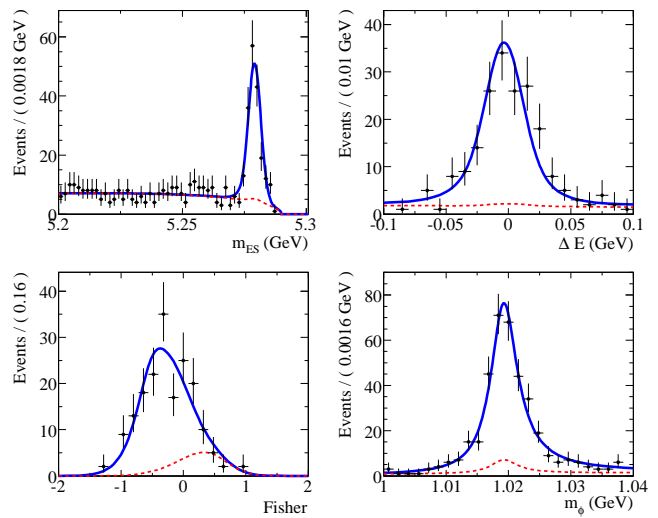}
       \caption{ Results of fitting the $B^+ \to \phi\phi K^+$ sample for
                 $m_{\phi\phi}<2.85$ GeV.
                 The dashed red curve is the sum of the combinatoric
                 and peaking background components.
                 The solid blue curve is for all components.
                 A requirement on a likelihood ratio based on all fit variables
                 except the one plotted is made to reject most of
                 the background.
                 The likelihood ratio requirements are about 84\% efficient for
                 the signal.
                 }
       \label{fig:bplus-fit}
     \end{center}
   \end{figure}

\begin{figure}[h!]
     \begin{center}
         \includegraphics[width=\linewidth]{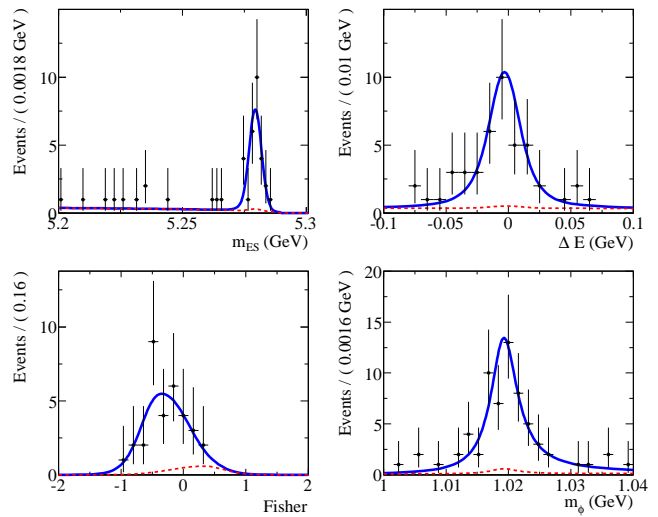}
       \caption{ Results of fitting the $B^0 \to \phi\phi K^0_S$ sample for
                 $m_{\phi\phi}<2.85$ GeV.
                 The dashed red curve is the sum of the combinatoric
                 and peaking background components.
                 The solid blue curve is for all components.
                 A requirement on a likelihood ratio based on all fit variables
                 except the one plotted is made to reject most of
                 the background.
                 The likelihood ratio requirements are about 84\% efficient for
                 the signal.
                 }
       \label{fig:b0-fit}
     \end{center}
   \end{figure}

   The results of the $B^+$ and $B^0$ fits are shown in Figs.~\ref{fig:bplus-fit}
   and~\ref{fig:b0-fit}, respectively.
   To reduce the combinatoric background in each distribution
   shown in the figures, a requirement is made on a likelihood
   ratio, which is based on all the fit variables except the one plotted.

   The fitted charge asymmetry ${\cal A}_{CP}$ for the background component
   is $0.02 \pm 0.03$.
   The charge asymmetry for the signal component is $-0.10 \pm 0.08$.
   The fitted yields of $B^+$ and $B^0$ signal candidates
   with $m_{\phi\phi}<2.85$ GeV
   are $178 \pm 15$ events and $40 \pm 7$ events, respectively,
   where the uncertainties are statistical only.


  \subsection{Systematic Uncertainties}

  Table~\ref{tab:bf-fit-syst} summarizes the systematic uncertainties on the
  $B\to \phi\phi K$ branching fractions in the $m_{\phi\phi}<2.85$ GeV region.
  The systematics are divided into additive uncertainties that affect the
  $B$ yield measurement
  and multiplicative uncertainties
  in the branching fraction calculation.

  The uncertainties from the corrections applied to the PDF parameters
  such as the $m_{\rm ES}$ and $\Delta E$ core mean and width
  for the signal component,
  which are derived from data control samples, are listed under ``ML Fit Yield''.
  The signal Fisher and $m_\phi$ core Gaussian mean and width parameters
  are not corrected in the fit, because data control sample measurements
  are consistent with the Monte Carlo.
  However, we did vary the signal Fisher and $m_\phi$ core Gaussian mean and
  width parameters by the statistical uncertainty of the data control sample
  measurements.  These variations are also included under ``ML Fit Yield''.
  The fit bias systematic is taken to be half of the bias correction added in
  quadrature with the statistical uncertainty on the bias.
  We vary the fixed peaking background yields by their statistical uncertainties
  (see Tables~\ref{tab:b+zone-fits} and~\ref{tab:b0zone-fits})
  and by varying the fractions $f_{ij}$.
  The fixed self-crossfeed fraction for the signal component was varied by $\pm 2$\%.
  Adding the individual uncertainties in quadrature, the total additive systematic
  uncertainties on the $B^+$ and $B^0$ signal yields are 6.2 and 1.8 events, respectively.

  The uncertainty on the track reconstruction efficiency is $\pm 0.23$\% per track,
  which is taken to be fully correlated for the charged kaons.
  The $K^0_S$ reconstruction efficiency has an uncertainty of 1.5\%.
  The $\phi\to K^+K^-$ and $K^0_S\to \pi^+\pi^-$ branching fractions are taken
  from the PDG~\cite{pdg} and are varied by their one standard deviation uncertainties.
  The systematic uncertainty on the $K^\pm$ identification criteria was estimated by
  comparing the ratio of the $B$ yield with the nominal selection to the $B$ yield
  requiring all $K^\pm$ to pass the tighter selection in the data and the MC samples.
  This gives an uncertainty of 3\% for the $B^+$ mode and 2\% for the $B^0$ mode.
  Adding the individual uncertainties in quadrature, the overall multiplicative
  systematic uncertainties are 3.6\% for the $B^+$ mode and 3.2\% for the $B^0$ mode.


   \begin{table}
     \begin{center}
       \caption{ Summary of the systematic and statistical uncertainties
          for the branching fraction measurements. }
       \label{tab:bf-fit-syst}
       \begin{tabular}{ccc}
          \hline \hline
          Quantity   & $B^+ \rightarrow \phi\phi K^+$ & $B^0 \rightarrow \phi\phi K^{0}$ \\
          Fit Stat. Uncertainty (events)  &  15.1  &  7.0  \\
          \hline
          \multicolumn{3}{c}{Additive Uncertainties (events)}                       \\
          ML Fit Yield           & 3.3     & 1.0  \\
          ML Fit Bias            & 1.6     & 0.2  \\
          Peaking BG, region fits     & 3.5     & 1.2  \\
          Peaking BG, $f_{ij}$ values  & 3.0     & 0.8  \\
          Self-Crossfeed Fraction           & 1.8     & 0.4  \\
          Total Additive Syst. (events)& 6.2     & 1.8  \\
          \hline
          \multicolumn{3}{c}{Multiplicative Uncertainties (\%)}  \\
           Tracking Efficiency    & 1.2 &  1.0     \\
           $K_s^0$ Reconstruction Efficiency   &  -  &  1.5     \\
           Number \BB     & 1.1 & 1.1      \\
           $\cal B$ ($\phi \rightarrow K^+ K^-$) & 1.2   & 1.2 \\
           $\cal B$ ($K_s^0 \rightarrow \pi^+ \pi^-$) & -  & 0.1 \\
           MC Statistics        & 0.1  & 0.1  \\
           $m_{\phi\phi}$  Cut Efficiency       & 0.2  & 0.3  \\
           $K^\pm$ Identification                 & 3.0  & 2.0  \\
          Total Multiplicative Syst. (\%)    & 3.6   & 3.2  \\
          \hline
          Total Systematic  [$\cal B$] ($\times 10^{-6}$) & 0.3    & 0.3 \\
          Statistical [$\cal B$] ($\times 10^{-6}$) & 0.5 & 0.8 \\
          \hline \hline
       \end{tabular}
     \end{center}
   \end{table}

   The signal charge asymmetry has been corrected for a bias due to differences in the
   $K^+$ and $K^-$ efficiencies by adding $+0.010\pm0.005$ to the asymmetry.
   The overall 2\% systematic uncertainty takes into account uncertainties
   on the charge dependence of the tracking efficiency, material interaction cross
   section for kaons, and particle identification.


  \subsection{Branching Fraction Results}

   Table~\ref{tab:bf-fit-results} summarizes the $B \to \phi\phi K$ branching
   fraction results for $m_{\phi\phi} < 2.85$ GeV.
   We find
   \begin{eqnarray*}
     {\cal B}(B^+ \to \phi\phi K^+) & = & (5.6 \pm 0.5 \pm 0.3) \times 10^{-6} \\
     {\cal B}(B^0 \to \phi\phi K^0) & = & (4.5 \pm 0.8 \pm 0.3) \times 10^{-6},
   \end{eqnarray*}
   where the first uncertainty is statistical and the second systematic.
   These results are consistent with and supersede the previous measurements~\cite{old-babar-ppk}
   by the \babar\ Collaboration.
   The Belle collaboration measurements~\cite{belle-ppk} are lower, though
   they are statistically compatible.
   Our branching fraction measurements are higher than the theoretical
   predictions of~\cite{fajfer} and~\cite{chen}.


   \begin{table}
     \begin{center}
       \caption{ Branching fraction and charge asymmetry results for $B\to \phi\phi K$ in the
           region $m_{\phi\phi} < 2.85$~GeV.
           The statistical significance is given by $\sqrt{2 \ln ({\cal L}_{\rm max}/{\cal L}_0)}$,
           where ${\cal L}_{\rm max}$ is the maximum likelihood and
           ${\cal L}_0$ is the likelihood for the hypothesis of no $\phi\phi K$ signal.
           The significance does not include systematic uncertainties.
         }
       \label{tab:bf-fit-results}
       \begin{tabular}{ccc}
          \hline \hline
                   & $B^+ \rightarrow \phi\phi K^+$ & $B^0 \rightarrow \phi\phi K^{0}$ \\
          \hline
          Events to fit              & 1535                        & 293                        \\
          Fit signal yield           & 178 $\pm$ 15                & 40 $\pm$ 7             \\
          ML-fit bias (events)       & 3.0 $\pm$ 0.5               & 0.0 $\pm$ 0.2              \\
          MC efficiency (\%)         & 28.0                        & 22.5                       \\
          $\Pi{\cal B}_i$(\%)        & 24.2                        & 8.4                        \\
          Stat. significance         & 24  &  11 \\
          \hline
          & & \\
          $\cal B$($10^{-6}$)        & 5.6 $\pm$ 0.5 $\pm$ 0.3  & 4.5 $\pm$ 0.8 $\pm$ 0.3  \\
          & & \\
          \hline
           Signal {\cal $A_{CP}$}     & $-0.10 \pm 0.08 \pm 0.02$               &  -                    \\
          Comb. Bkg. {\cal $A_{CP}$} & 0.02 $\pm$ 0.03                         &  -                    \\
          \hline \hline
       \end{tabular}
     \end{center}
   \end{table}


  \section{\boldmath CP asymmetry in $\eta_c$ resonance region}

\begin{figure}
     \begin{center}
         \includegraphics[width=\linewidth]{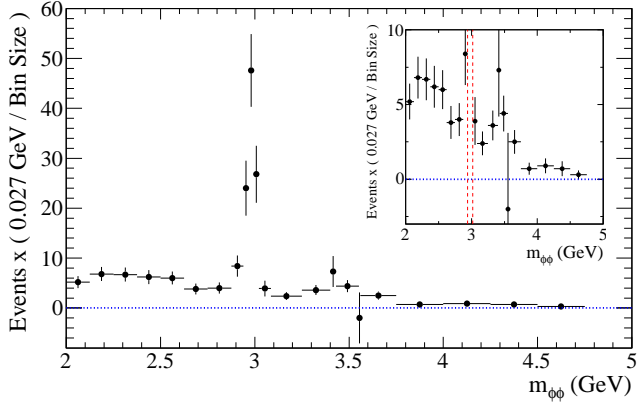}
       \caption{ Fitted $B^+\to \phi\phi K^+$ yield as a function
                  of $m_{\phi\phi}$.
                  Each point shows the results of a maximum likelihood
                  fit of the events in that bin.
                  The inset is the same data with an expanded vertical
                  range to show the shape of the non-resonant component
                  more clearly.
                  The yield has been divided by the bin width
                  and scaled by 0.027 GeV, which is the bin width
                  of the three bins in the
                  $\eta_c$ resonance region ([2.94,3.02] GeV
                  and dashed vertical lines in the inset).
                  The two narrow bins above the $\eta_c$ are
                  centered on the $\chi_{c0}$ (bin range [3.400,3.430] GeV)
                  and the $\chi_{c2}$ (bin range [3.552,3.560] GeV).
                 }
       \label{fig:mphiphi}
     \end{center}
   \end{figure}

   As was mentioned in the introduction, a significant non-zero direct $CP$ asymmetry
   in the $\eta_c$ resonance region of $m_{\phi\phi}$ would be a clear
   sign of physics beyond the Standard Model.
   For this measurement, we use the simpler likelihood, based on
   $m_{\rm ES}$, $\Delta E$, and ${\cal F}$.
   Figure~\ref{fig:mphiphi} shows the fitted $B^+\to \phi\phi K^+$ yield as a function
   of $m_{\phi\phi}$.
   The $\eta_c$ resonance is clearly visible.
   Narrow bins around the $\chi_{c0}$ and $\chi_{c2}$ resonances do not show a
   significant excess above the broad non-resonant component.

   The results of fitting the events in the $m_{\phi\phi}$ range
   of [2.94,3.02] GeV are given in Table~\ref{tab:etac-region-fit-results}.
   For the $CP$ asymmetry, we find
   \begin{equation*}
      A_{CP}(m_{\phi\phi}\ {\rm in} \ [2.94,3.02] \ {\rm GeV}) = -0.09 \pm 0.10 \pm 0.02 ,
   \end{equation*}
   where the first uncertainty is statistical and the second uncertainty is
   systematic.
   The value above includes the same 1\% bias correction
   and has the same 2\% overall systematic uncertainty as
   the signal charge asymmetry below the $\eta_c$ resonance
   as described above.

   The fit yields $100 \pm 10$ signal candidates.
   Using ${\cal B}(B^+\to \eta_c K^+) = (9.1 \pm 1.3) \times 10^{-4}$
   and ${\cal B}(\eta_c \to \phi\phi) = (2.7 \pm 0.9) \times 10^{-3}$ from the PDG~\cite{pdg},
   a $B^+\to \phi\phi K^+$; $\phi\to K^+K^-$ reconstruction efficiency of 29\%
   in the $\eta_c$ resonance region,
   and an efficiency of 78\% for the $m_{\phi\phi}$ window of
   [2.94,3.02] GeV for the $\eta_c$ resonance, we would
   expect $62 \pm 22$ signal events, ignoring the non-resonant $B^+\to \phi\phi K^+$
   contribution and any interference between the resonant $\eta_c$ and non-resonant
   amplitudes.
   We do not use our $B^+$ event yield to measure
   ${\cal B}(B^+\to \eta_c K^+) \times {\cal B}(\eta_c \to \phi\phi)$
   due to
   the potentially large interference effects between the resonant and non-resonant
   $\phi\phi$ amplitudes which we can not easily quantify.

   The $A_{CP}$ may integrate to zero, even if there is a contributing
   non-Standard-Model amplitude with a non-zero $CP$ violating phase.
   However, in this case the phase variation of the $\eta_c$ resonance amplitude
   could give non-zero $A_{CP}$ values with opposite signs
   above and below the peak of the resonance.
   We have performed the measurement in two ranges, splitting the $\eta_c$
   region into two regions (above and below the peak of the resonance).
   The results are
   \begin{eqnarray*}
      A_{CP}(m_{\phi\phi}\ {\rm in} \ [2.94,2.98] \ {\rm GeV}) & = & -0.10 \pm 0.15 \pm 0.02 \\
      A_{CP}(m_{\phi\phi}\ {\rm in} \ [2.98,3.02] \ {\rm GeV}) & = & -0.08 \pm 0.14 \pm 0.02,
   \end{eqnarray*}
   both of which are consistent with zero, as expected in the Standard Model.


    \begin{table}
     \begin{center}
     \caption{Fit results for $B^+\to \phi\phi K^+$ within $\eta_c$ 
           resonance region ($m_{\phi\phi}$ within [2.94,3.02] GeV).
           The signal charge asymmetry {\cal $A_{CP}$} has been
           corrected by adding $+0.010 \pm 0.005$ to the fitted asymmetry.
             }
     \label{tab:etac-region-fit-results}
     \begin{tabular}{cc}
     \hline \hline
     ML fit quantity/Analysis   & $B^+ \rightarrow \phi\phi K^+$  \\
     \hline
     Events to fit              & 181                             \\
     Fit signal yield           & 100 $\pm$ 10                \\
     MC efficiency (\%)         & 29.2                             \\
     \hline
     Corr. Signal {\cal $A_{CP}$}     & $-0.09 \pm 0.10 \pm 0.02$ \\
     Comb. Bkg. {\cal $A_{CP}$} & $-0.06 \pm 0.11$        \\
     \hline \hline
     \end{tabular}
     \end{center}
    \end{table}


  \section{Angular Studies}

   We use the angular variables that describe the $B^+\to \phi\phi K^+$ decay
   to investigate the spin components of the $\phi\phi$ system
   below and within the $\eta_c$ resonance.
   The angles are defined as follows.
   \begin{itemize}
     \item {\boldmath $\theta_i$}, $(i=1,2)$ : The $\theta_i$ angle is
          the angle between the momentum of the $K^+$ coming from the
          decay of $\phi_i$
          in the $\phi_i$ rest frame with respect to the boost direction from
          the $\phi\phi$ rest frame to the $\phi_i$ rest frame.
     \item {\boldmath $\chi$}: The $\chi$ angle is the dihedral angle between
          the $\phi_1$ and $\phi_2$ decay planes in the $\phi\phi$ rest frame.
     \item {\boldmath $\theta_{\phi\phi}$}: The $\theta_{\phi\phi}$ angle
          is the angle between one of the $\phi$ mesons in $\phi\phi$ rest
          frame with respect to the boost direction from the $B^+$ rest frame
          to the $\phi\phi$ rest frame.
   \end{itemize}

   We project the $J^P = 0^-$ component by making a histogram
   of $m_{\phi\phi}$ weighting each event by
   \begin{equation}
   \begin{split}
           P_2(\cos \theta_1) \ {\rm Re} \left[ Y^2_2( \theta_2, \chi) \right] \ = \\
                   \frac{25}{4} \left\{ 3 \cos^2 \theta_1  - 1 \right\}
                            \, \sin^2 \theta_2  \, \cos  2 \chi,
   \end{split}
   \end{equation}
   where $P_2$ is a second-degree Legendre polynomial and $Y^2_2$ is
   a spherical harmonic with $\ell=2$ and $m=2$.
   In each bin, the $J^P = 0^-$ component yield is projected out,
   while the combinatoric background averages to zero.
   To do this, we select events in a signal region defined by:
   $m_{\rm ES} > 5.27$~GeV, $|\Delta E|< 40$~MeV, $m_\phi$ within [1.01,1.03]~GeV,
   and 
   ${\cal F}<0.5$.
   The efficiency of these requirements, relative to the selection used
   in the asymmetry measurement, is about 78\% for signal events and 2.9\%
   for combinatoric background.
   The combinatoric background that remains after this selection is shown using
   data events in the sideband region ($m_{\rm ES} <5.27$~GeV and $|\Delta E|<100$~MeV)
   scaled by 0.065, which is the signal-to-sideband ratio for the combinatoric
   background.

\begin{figure}
     \begin{center}
         \includegraphics[width=0.99\linewidth]{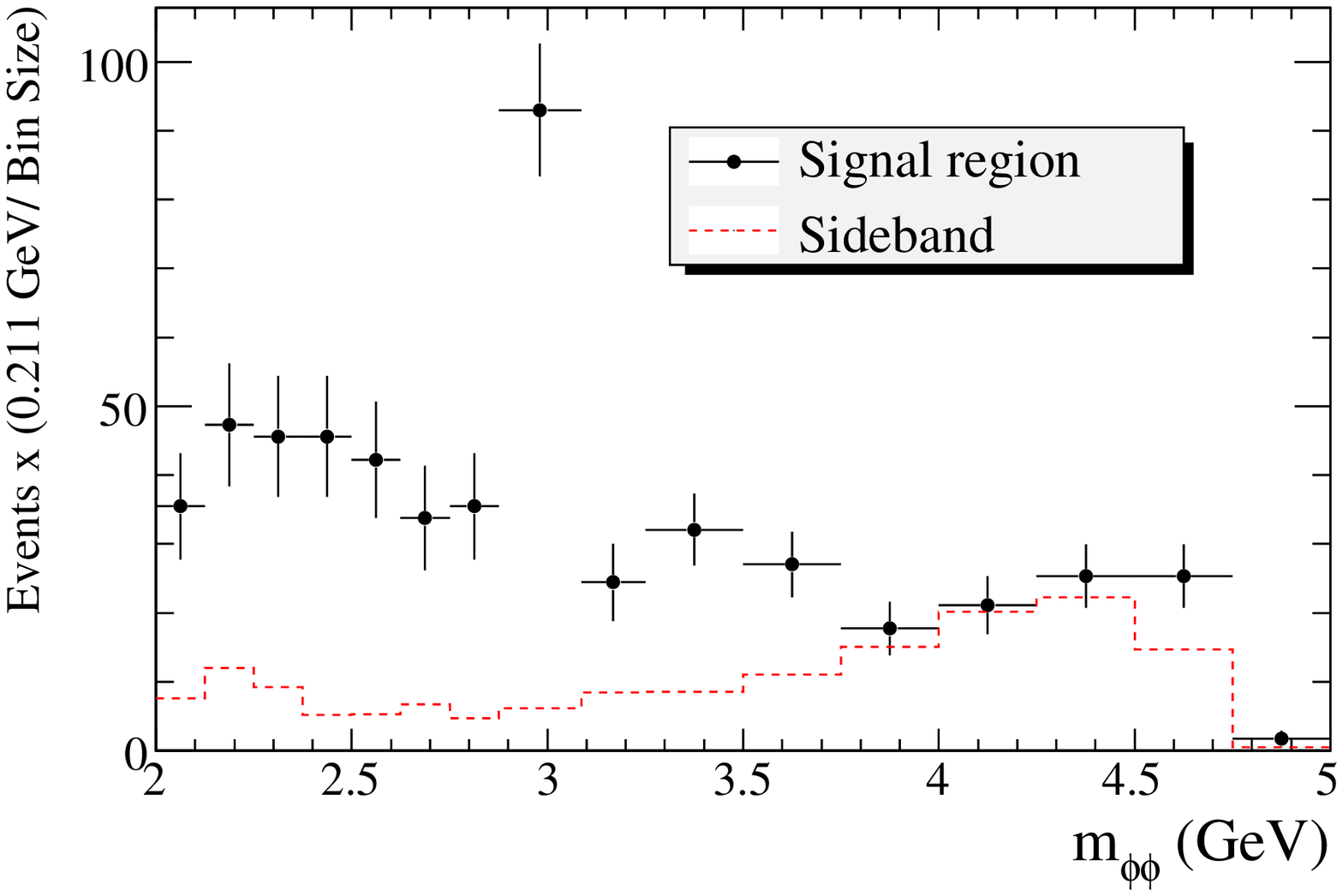}
         \includegraphics[width=0.99\linewidth]{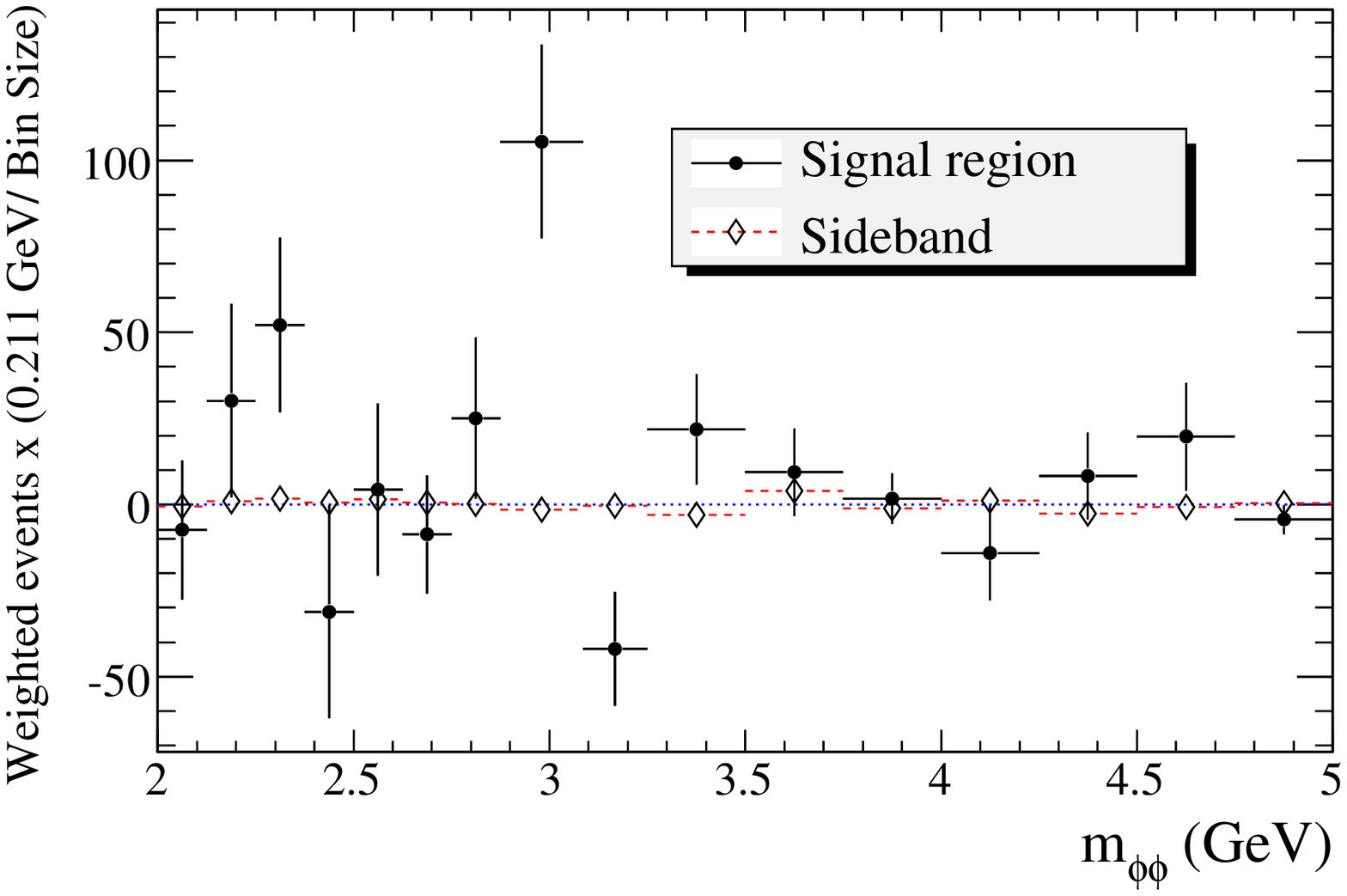}
       \caption{
           Histograms (top) and weighted distributions (bottom) of $m_{\phi\phi}$ for the
           signal region (solid points) and data sideband selection (dashed with
           open diamonds) defined in the text.
           The sideband distributions have been normalized to the expected level
           of combinatoric background remaining after the signal region selection.
           Events in the bottom distribution were weighted by
           $P_2(\cos \theta_1) \ {\rm Re} \left[ Y^2_2( \theta_2, \chi) \right]$
           which projects out the $J^P = 0^-$ component.
                  The yield has been divided by the bin width
                  and scaled by 0.211 GeV, which is the
                  width of the  bin covering the
                  $\eta_c$ resonance ([2.875,3.086] GeV).
                 }
       \label{fig:angular-weight}
     \end{center}
   \end{figure}

   The results are shown in Fig.~\ref{fig:angular-weight}.
   The weighted yield in the $\eta_c$ region is consistent with all of the
   $B^+\to \phi\phi K^+$ events having $J^P = 0^-$.
   Just below the $\eta_c$ region, the weighted yield is consistent with zero.
   The excess in the bins near 2.2~GeV may be due
   to the $\eta(2225)$ seen in $J/\psi \to \gamma \phi\phi$ events at
   Mark III~\cite{markIII} and BES~\cite{BES}.

\begin{figure*}
     \begin{center}
         \includegraphics[width=0.32\linewidth]{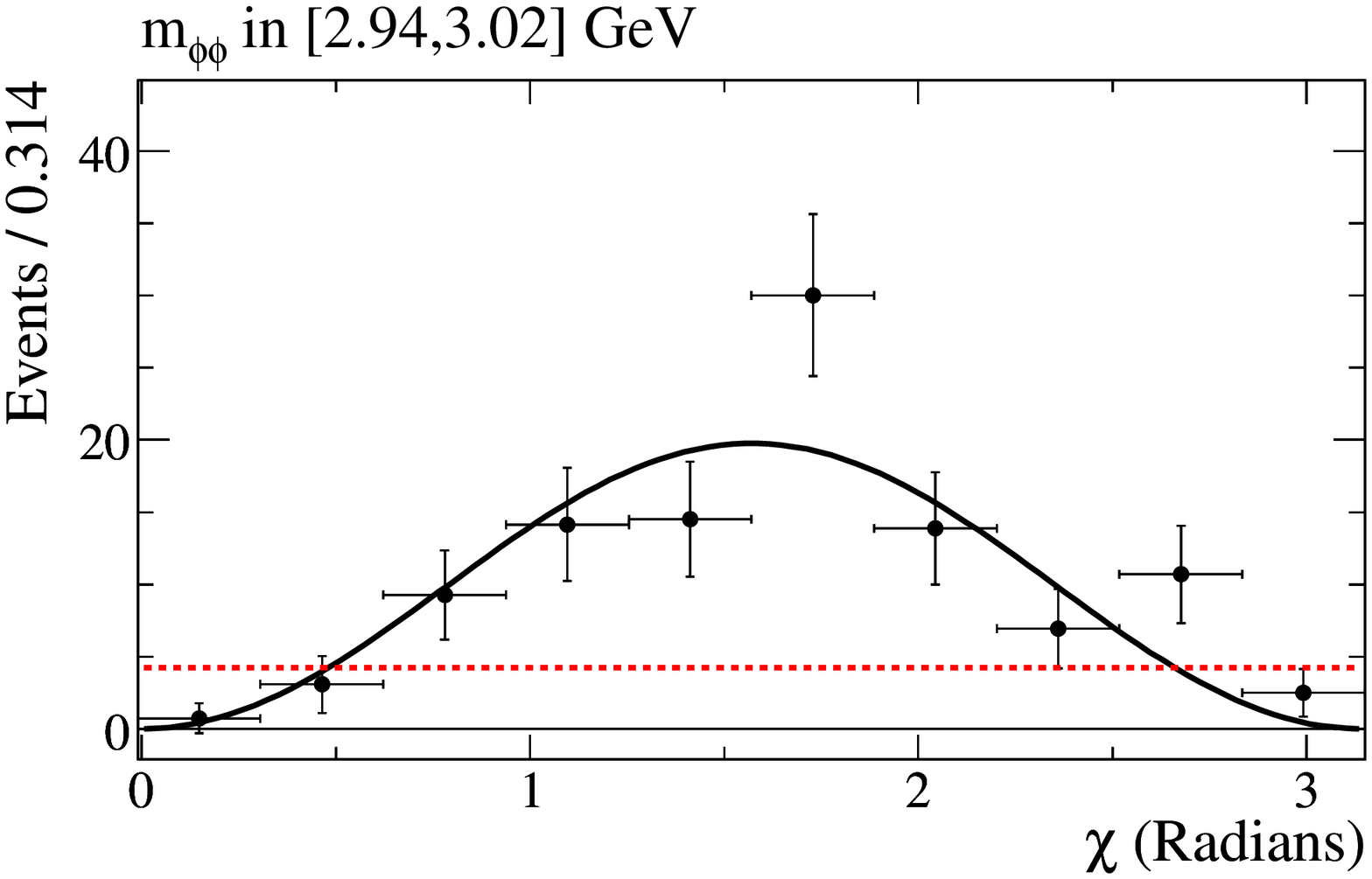}
         \includegraphics[width=0.32\linewidth]{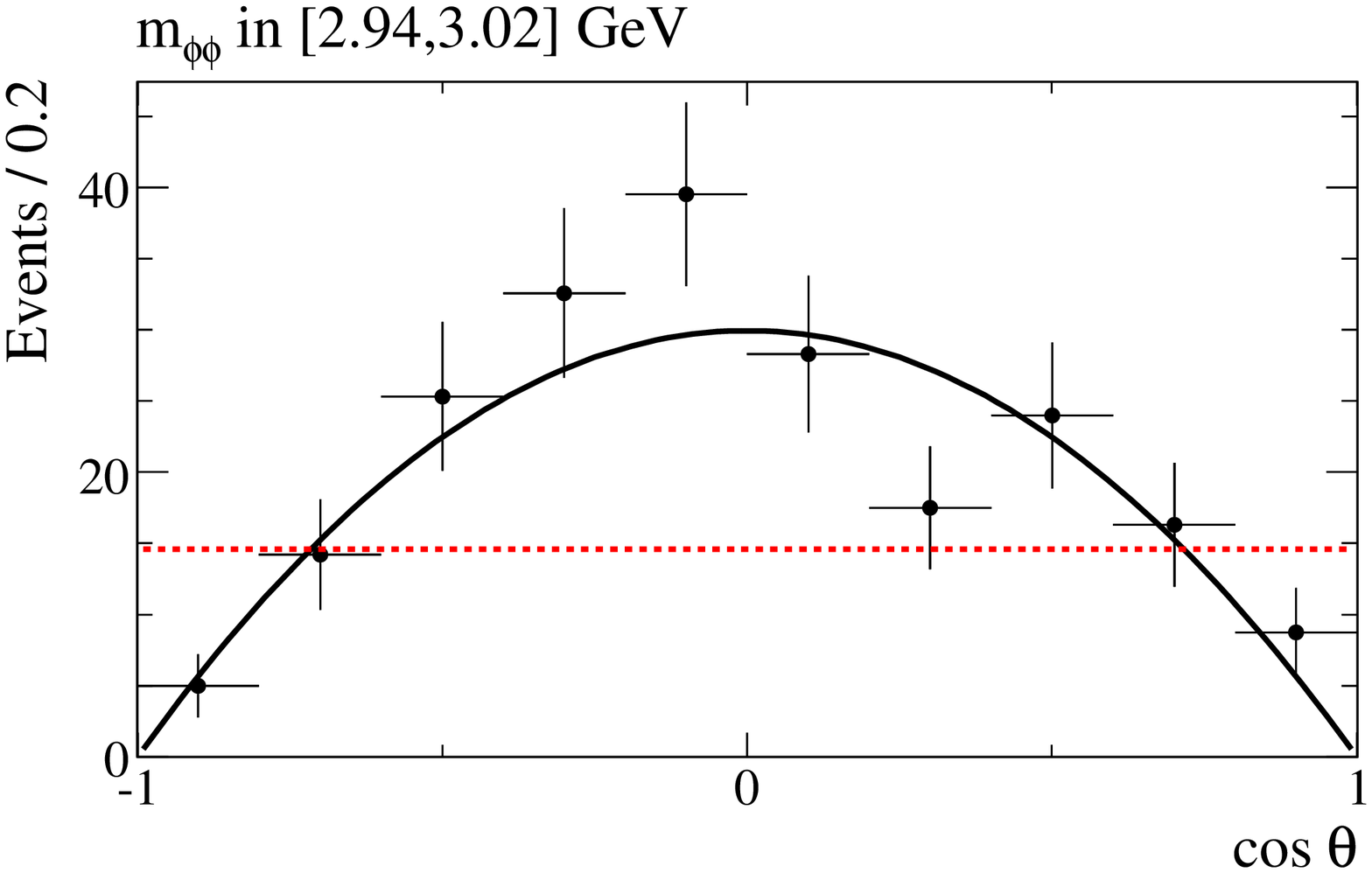}
         \includegraphics[width=0.32\linewidth]{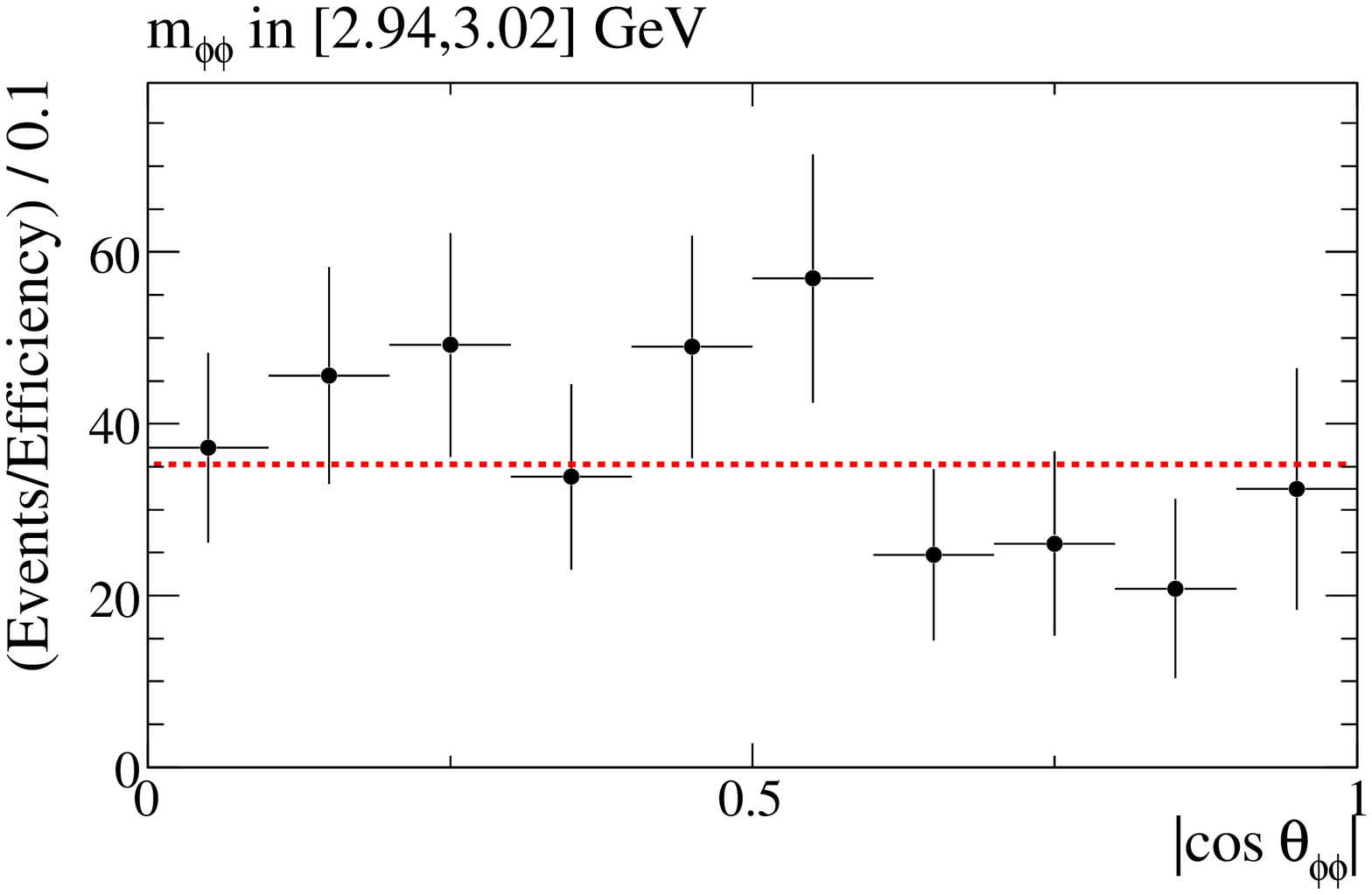}
         \includegraphics[width=0.32\linewidth]{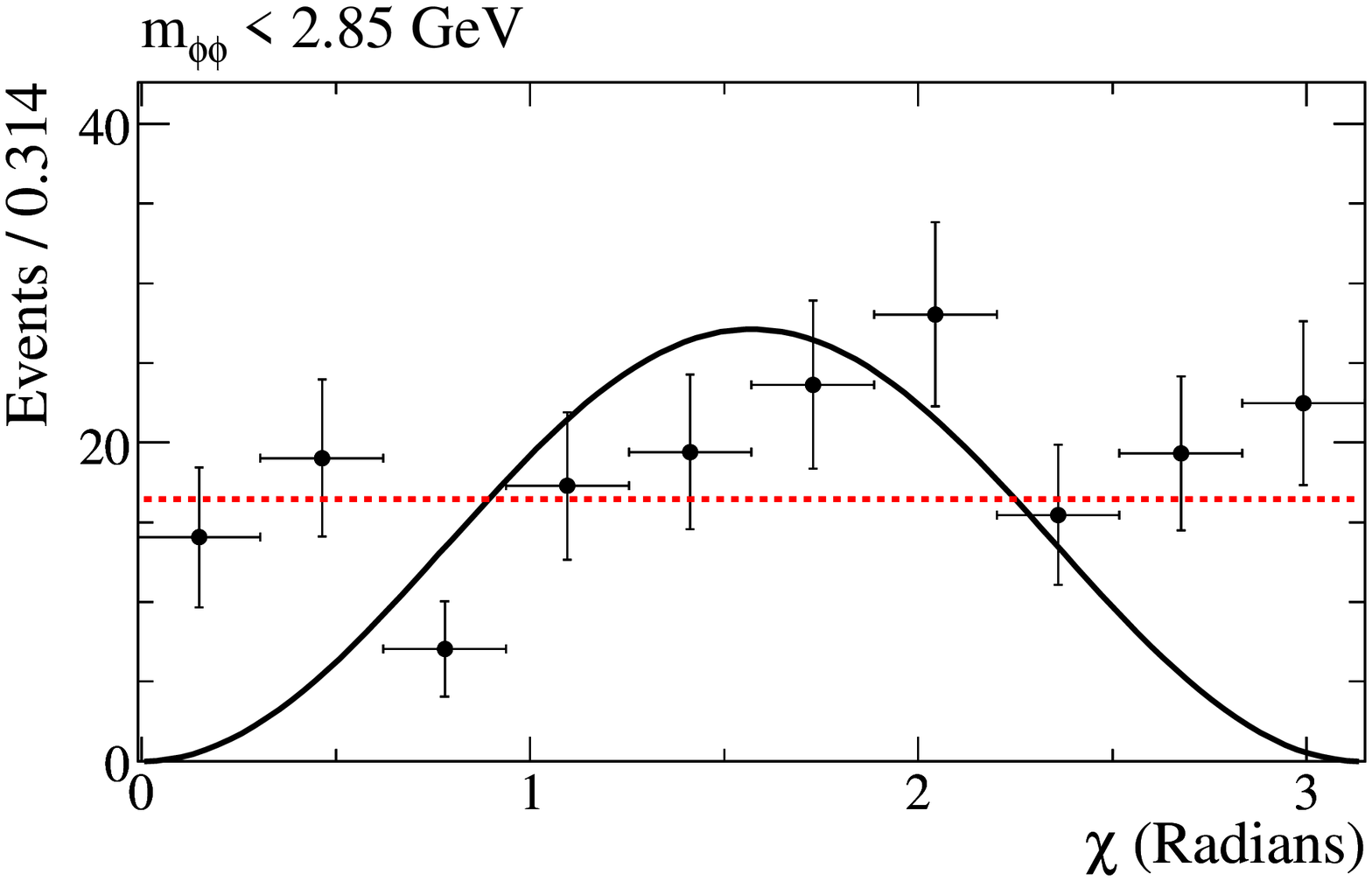}
         \includegraphics[width=0.32\linewidth]{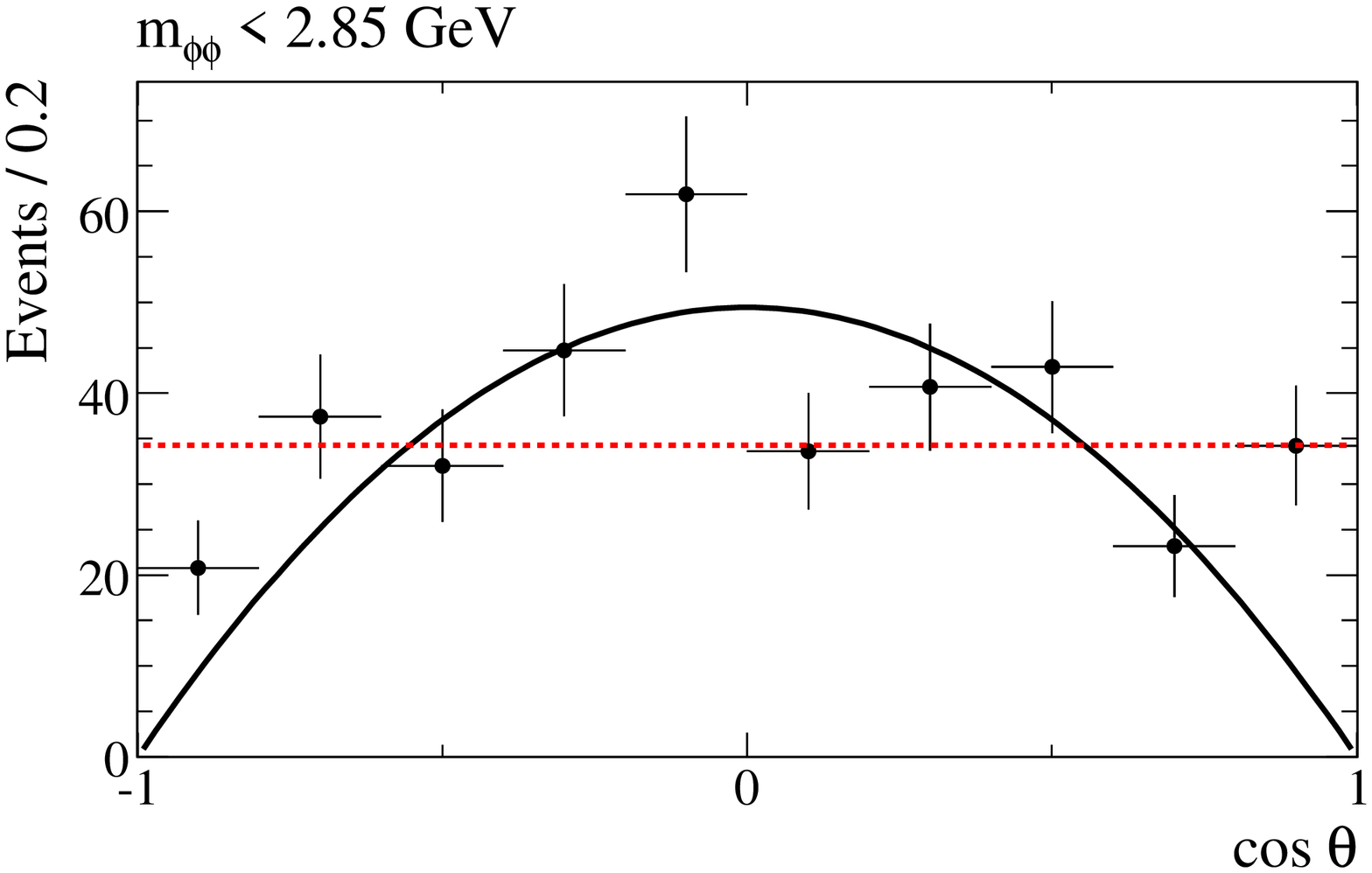}
         \includegraphics[width=0.32\linewidth]{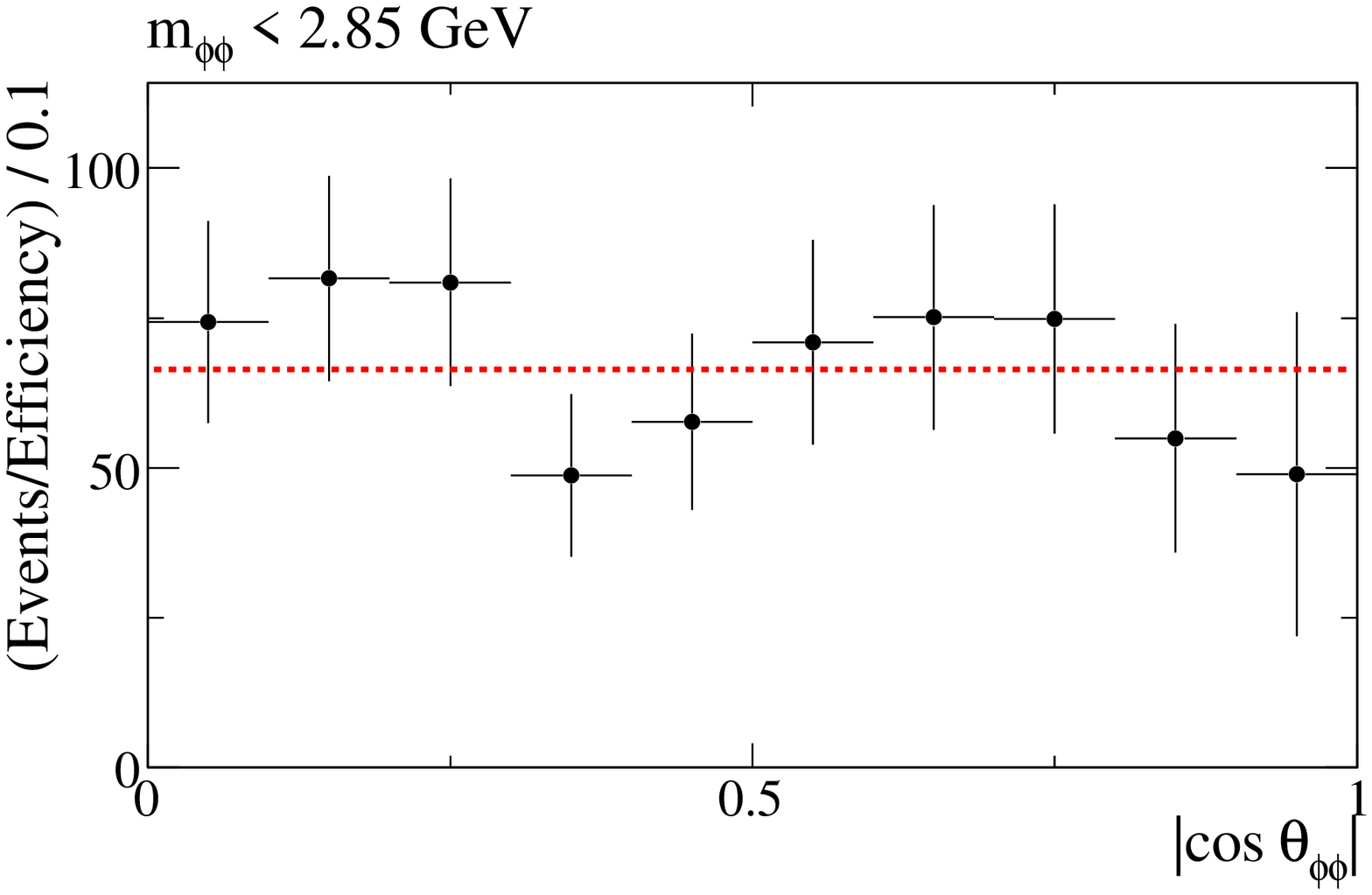}
       \caption{
          Background subtracted angular distributions in the $\eta_c$ resonance
            region ($m_{\phi\phi}$ in [2.94,3.02] GeV for the top row) and below
            the $\eta_c$ resonance ($m_{\phi\phi} < 2.85$ GeV for the bottom row).
            The reconstruction and selection efficiency is uniform in
            $\chi$ (left) and $\cos \theta $ (center), but dependent on
            $|\cos \theta_{\phi\phi} |$ (right), so the right column has
            been efficiency corrected.
            The red dashed line shows a least-$\chi^2$ fit of the points to a uniform
            distribution while the solid black curve shows a fit to the expectation
            for a $J^P = 0^-$ state decaying to $\phi\phi$.
                 }
       \label{fig:angular-plots}
     \end{center}
   \end{figure*}

   Figure~\ref{fig:angular-plots} shows background-subtracted distributions
   of $\chi$, $\cos\theta_i$, and $|\cos\theta_{\phi\phi}|$ for the
   nominal event selection.
   The background subtraction is done with the technique described
   in reference~\cite{splots}.
   Since there is no meaningful distinction between $\phi_1$ and $\phi_2$,
   we combine the $\cos\theta_1$ and $\cos\theta_2$ distributions into
   one plot of $\cos\theta$.
   The reconstruction and selection efficiency, determined
   from MC samples, is uniform in $\chi$ and
   $\cos\theta_1$, but not in $|\cos\theta_{\phi\phi}|$, so the
   $|\cos\theta_{\phi\phi}|$ distribution is efficiency corrected.
   For each distribution, we performed a simple least-$\chi^2$ fit
   to the distributions expected
   for both  $J^P = 0^-$  and $J^P = 0^+$ for the $\phi\phi$ system.

   \begin{table}
     \begin{center}
       \caption{Quality of the angular fits shown in Fig.~\ref{fig:angular-plots}.
                The first column is the $m_{\phi\phi}$ interval for the events in the
                fit.  The last column is the $p$-value of the $\chi^2$ goodness-of-fit
                test for the hypothesis indicated in the third column.}
       \label{tab:angular-fits}
       \begin{tabular}{ccccc}
       \hline\hline
              $m_{\phi\phi}$ (GeV)
          &   \ \ Variable \ \
          &   PDF
          &   $\chi^2/N_{\rm dof}$
          &   $\chi^2$ prob.  \\
         \hline
          $[2.94, 3.02]$  &   $\chi$   &   $\sin^2 \chi$   & \ \ \ $9.51/9$ \ \ \ &   $0.39$   \\
          $[2.94, 3.02]$  &   $\chi$   &   uniform         &   $60.3/9$   &   $1.2\times 10^{-9}$   \\
          $<2.85$         &   $\chi$   &   $\sin^2 \chi$   &   $41.6/9$   &   $3.9\times 10^{-6}$ \\
          $<2.85$         &   $\chi$   &   uniform         &   $18.5/9$   &   $0.030$ \\
         \hline
          $[2.94, 3.02]$  &   $\cos \theta$   &   $\sin^2 \chi$   &   $9.97/9$   &   $0.39$   \\
          $[2.94, 3.02]$  &   $\cos \theta$   &   uniform         &   $60.5/9$   &   $1.1\times 10^{-9}$   \\
          $<2.85$         &   $\cos \theta$   &   $\sin^2 \chi$   &   $32.9/9$   &   $1.7\times 10^{-4}$ \\
          $<2.85$         &   $\cos \theta$   &   uniform         &   $25.8/9$   &   $2.2\times 10^{-3}$ \\
         \hline
          $[2.94, 3.02]$  &   $|\cos \theta_{\phi\phi}|$   &   uniform   &   $9.02/9$   &   $0.44$   \\
          $<2.85$         &   $|\cos \theta_{\phi\phi}|$   &   uniform   &   $5.01/9$   &   $0.83$ \\
         \hline
       \end{tabular}
     \end{center}
   \end{table}


   For a $J^P = 0^-$ state, we expect $\chi$ to have a $\sin^2 \chi$
   distribution, while $\chi$ should be uniform for $J^P = 0^+$.
   The signal events in the $\eta_c$ resonance region are consistent with a
   $\sin^2 \chi$ distribution while the signal below the $\eta_c$ resonance
   is not.
   For a $J^P = 0^-$ state, the distributions of $\cos \theta_i$ are
   expected to have $\sin^2 \theta_i$ distributions, while
   a $J^P = 0^+$ state is expected to have uniform $\cos \theta_i$ distributions.
   The events in the $\eta_c$ resonance region are consistent with
   a $\sin^2 \theta_i$ distribution,
   while the events below the $\eta_c$ resonance show a deviation from
   a $\sin^2 \theta_i$ shape.

   Finally, a spin-zero state should have a uniform $|\cos \theta_{\phi\phi} |$
   distribution.
   The efficiency-corrected distributions shown in
   Fig.~\ref{fig:angular-plots}, both within and below the $\eta_c$
   resonance region, are consistent with a uniform $|\cos \theta_{\phi\phi} |$
   distribution.


  \section{Summary and Conclusions}

   We have measured the branching fractions and charge asymmetries of
   $B\to \phi\phi K$ decays below the $\eta_c$ resonance in the
   $\phi\phi$ invariant mass ($m_{\phi\phi} < 2.85$ GeV).
   We observe both $B^+ \to \phi\phi K^+$ and $B^0 \to \phi\phi K^0_S$,
   each with a significance of greater than five standard deviations.
   The $B^0 \to \phi\phi K^0_S$ decay has not been observed previously.
   Our branching fraction measurements are higher than the theoretical
   predictions of~\cite{fajfer} and~\cite{chen}.

   We have measured the charge asymmetry for $B^+\to \phi\phi K^+$ in the
   $\eta_c$ resonance region, where a significant non-zero value would
   be an unambiguous indication of new physics.
   Our measurement is consistent with zero, which is the expectation of
   the Standard Model.

   Finally, we have studied the angular distributions of $B^+ \to \phi\phi K^+$
   decays below and within the $\eta_c$ resonance.
   We conclude from these studies that the non-resonant $B^+ \to \phi\phi K^+$
   events below the $\eta_c$ resonance are, on average, more consistent with
   $J^P = 0^+$ than $J^P = 0^-$, while
   the distributions within the $\eta_c$ resonance region are all consistent
   with $J^P = 0^-$.

   \par
   We are grateful for the 
extraordinary contributions of our \pep2\ colleagues in
achieving the excellent luminosity and machine conditions
that have made this work possible.
The success of this project also relies critically on the 
expertise and dedication of the computing organizations that 
support \babar.
The collaborating institutions wish to thank 
SLAC for its support and the kind hospitality extended to them. 
This work is supported by the
US Department of Energy
and National Science Foundation, the
Natural Sciences and Engineering Research Council (Canada),
the Commissariat \`a l'Energie Atomique and
Institut National de Physique Nucl\'eaire et de Physique des Particules
(France), the
Bundesministerium f\"ur Bildung und Forschung and
Deutsche Forschungsgemeinschaft
(Germany), the
Istituto Nazionale di Fisica Nucleare (Italy),
the Foundation for Fundamental Research on Matter (The Netherlands),
the Research Council of Norway, the
Ministry of Education and Science of the Russian Federation, 
Ministerio de Ciencia e Innovaci\'on (Spain), and the
Science and Technology Facilities Council (United Kingdom).
Individuals have received support from 
the Marie-Curie IEF program (European Union), the A. P. Sloan Foundation (USA) 
and the Binational Science Foundation (USA-Israel).


 \appendix

 \vspace{1cm}
\section{ Peaking background zone fit results}

   The results of the fits of the $B\to 5K$ yield in
   the five zones in the $m_{\phi 2}$ vs $m_{\phi 1}$
   plane (shown in Fig.~\ref{fig:mphi1vsmphi2})
   and the derived yields for each of the five $B$ decay
   modes are given in Tables~\ref{tab:b+zone-fits} and~~\ref{tab:b0zone-fits}
   below.

   \begin{table*}
    \begin{center}
     \caption{
              $B^+ \to 5K$ yield fit results for the five zones ($B^+$ yield column)
               and yields derived from the fraction matrix $f_{ij}$.
              The last row is computed from the inverted fraction matrix
              $f_{ij}^{-1}$ and the $B^+$ yield column.
              The remaining $5\times 5$ yield matrix was computed from the last row and the
              fraction matrix $f_{ij}$.
              The uncertainties are from propagating the statistical errors from
              the fitted region yields ($B^+$ yield column) without including any
              uncertainties on the fraction matrix.
              }
     \label{tab:b+zone-fits}
     \begin{tabular}{ccccccc}
     \hline
     \hspace{4mm} Zone \hspace{4mm}  &  \hspace{4mm} $B^+$ yield  \hspace{4mm} & \ppkpm  &  \pkkkpm  & \kkkkkpm  &  \fzpkpm  &  \fzkkkpm  \\
     \hline \hline
     1        & \hspace{4mm} $188.4 \pm 16.0$ \hspace{4mm} & \hspace{4mm} $183.2 \pm 17.0$ \hspace{4mm}  & \hspace{4mm} $6.3 \pm 3.9$ \hspace{4mm}     & \hspace{4mm} $0.8 \pm 0.8$ \hspace{4mm}  &  \hspace{4mm} $0 \pm 3.0$ \hspace{4mm}  & \hspace{4mm}  $0.3 \pm 1.5$ \hspace{4mm}  \\
     \hline
     2        & \hspace{4mm} $ 84.4 \pm 18.0$ \hspace{4mm} &  36  &  39  &  9  &  -1.4  &  2  \\
     \hline
     3        & \hspace{4mm} $ 49.7 \pm 19.0$ \hspace{4mm} &   3.8  &  18  &  26  &  -0.2  &  2.2  \\
     \hline
     4        & \hspace{4mm} $  1.0 \pm  2.0$ \hspace{4mm} &   1.3  &  0.9  &  0.2  &  -1.7  &  0.3  \\
     \hline
     5        & \hspace{4mm} $  3.5 \pm  5.0$ \hspace{4mm} &   0.2  &  0.8  &  1.3  &  -0.2  &  1.3  \\
     \hline \hline
      $1-5$  & & $225 \pm 21$  &  \hspace{4mm} $65 \pm 40$ \hspace{4mm} &   $38 \pm 38$ &  \hspace{4mm} $-5.7 \pm 7.6$ \hspace{4mm}  &  \hspace{4mm} $6.2 \pm 26$ \hspace{4mm} \\
     \hline
     \end{tabular}
    \end{center}
   \end{table*}
   \begin{table*}
    \begin{center}
     \caption{
              $B^0 \to 5K$ yield fit results for the five zones ($B^0$ yield column)
               and yields derived from the fraction matrix $f_{ij}$.
              The last row is computed from the inverted fraction matrix
              $f_{ij}^{-1}$ and the $B^0$ yield column.
              The remaining $5\times 5$ yield matrix was computed from the last row and the
              fraction matrix $f_{ij}$.
              The uncertainties are from propagating the statistical errors from
              the fitted region yields ($B^0$ yield column) without including any
              uncertainties on the fraction matrix.
              }
     \label{tab:b0zone-fits}
     \begin{tabular}{ccccccc}
     \hline
     \hspace{4mm} Zone \hspace{4mm} & $B^0$ yield & \ppks  &  \pkkks  & \kkkkks  &  \fzpks  &  \fzkkks  \\
     \hline \hline
     1        & \hspace{4mm} $43.4 \pm 8.0$ \hspace{4mm}   & \hspace{4mm} $42.8 \pm 8.4$ \hspace{4mm}  & \hspace{4mm} $1.2 \pm 1.5$\hspace{4mm}   & \hspace{4mm} $0.1 \pm 0.3$ \hspace{4mm} & \hspace{4mm} $0 \pm 0.7$ \hspace{4mm}  & \hspace{4mm} $0 \pm 0.02$ \hspace{4mm} \\
     \hline
     2        & $15.1 \pm 7.0$   & 8      & 7    & 0.9  &  -0.4  &  -0.4  \\
     \hline
     3        & $ 6.0 \pm 6.0$   & 1      & 3    & 2.5  &  -0.1  & -0.4 \\
     \hline
     4        & $ 0 \pm 0.5$   & 0.3      & 0.2    & 0  &  -0.5  &  0 \\
     \hline
     5        & $ 0 \pm 0.8$   & 0      & 0.1    & 0.1  &  0  &  -0.2 \\
     \hline \hline
     $1-5$ & & $52 \pm 10$  &  $12 \pm 15$  &  $3.6 \pm 12$ & \hspace{4mm}  $-1.6 \pm 1.8$ \hspace{4mm}  & \hspace{4mm}  $-1.1 \pm 4.3$ \hspace{4mm} \\
     \hline
     \end{tabular}
    \end{center}
   \end{table*}



\begin{thebibliography}{99}
    %
    \bibitem{sakarov} A.D. Sakharov, JETP Lett. {\bf 5}, 24 (1967).
    %
    \bibitem{babarnim} B. Aubert {\it et al.} [The \babar\ Collaboration],
         Nucl. Instrum. Meth. {\bf A479}, 1 (2002).
    %
    \bibitem{bellenim} A. Abashian {\it et al.} [The Belle Collaboration],
         Nucl. Instrum. Meth. {\bf A479}, 117 (2002).
    %
    \bibitem{pepii} PEP-II Conceptual Design Report, SLAC-0418 (1993).
    %
    \bibitem{kekb} S. Kurokawa and E. Kikutani,
         Nucl. Instrum. Meth. {\bf A499}, 1 (2003).
    %
    \bibitem{km} M. Kobayashi and T. Maskawa, Prog. Theor. Phys. {\bf 49}, 652 (1973).
    %
    \bibitem{matter-antimatter-asym} V. A. Rubakov and M. E. Shaposhnikov, Phys. Usp. {\bf 39}, 461 (1996).
    %
    \bibitem{grossman} Y. Grossman and M. P. Worah, Phys. Lett. {\bf B 395}, 241 (1997).
    %
    \bibitem{hazumi} M. Hazumi, Phys. Lett. {\bf B 583}, 285 (2004).
    %
    \bibitem{belle-ppk} H. C. Huang {\it et al.} [The Belle Collaboration],
            Phys. Rev. Lett. {\bf 91}, 241802 (2003).;
            Y. T. Shen, K. F. Chen, P. Chang {\it et al.} [The Belle Collaboration],
            arXiv:0802.1547[hep-ex].
    %
    \bibitem{old-babar-ppk} B. Aubert {\it et al.} [The \babar\ Collaboration], Phys. Rev. Lett. {\bf 97}, 261803 (2006).
    %
    \bibitem{fajfer} S. Fajfer, T. N. Pham, and A. Prapotnik,
         Phys. Rev. {\bf D69}, 114020 (2004).
    %
    \bibitem{chen} C.-H. Chen and H.-n. Li,
         Phys. Rev. {\bf D70}, 054006 (2004).
    %
    \bibitem{chargeconj} Charge-conjugate states are implicitly included throughout the
       paper unless stated otherwise.
    %
    \bibitem{c=1} Throughout the paper, necessary factors of $c$ are implied
        in the units for energy, mass, and momentum.
    %
    \bibitem{evtgen} D. Lange, Nucl. Instrum. Meth., A {\bf 462}, 152 (2001).
    %
    \bibitem{geant4} S. Agostinelli {\it et al.} [The \textsc{Geant4} Collaboration], Nucl. Instrum. Meth., A {\bf 506}, 250 (2003).
    %
    \bibitem{fox-wolfram} G. Fox and S. Wolfram, Phys. Rev. Lett., {\bf 41}, 1581 (1978).
    %
    \bibitem{dtandft} B. Aubert {\it et al.} [The \babar\ Collaboration],
          Phys. Rev. {\bf D79}, 072009 (2009).
    %
    \bibitem{phi-assignment}  
      In each event one of the $\phi$ candidates is randomly chosen to
      be $\phi_1$ with the other $\phi_2$.
      All aspects of the analysis, such as the PDFs used in the
      likelihood fits, are symmetric under the exchange of the $\phi_1$ and $\phi_2$ assignments
      (e.g. $m_{\phi 1} \leftrightarrow m_{\phi 2}$).
   %
    %
    \bibitem{CB-function} J. E. Gaiser, Ph.D. thesis, Stanford University
      [SLAC-R-255] (1982).
    %
    \bibitem{Argus-function}  With $x\equiv m_{\rm ES}/E_{\rm beam}^{*}$ and $\xi$ a
       parameter to be fitted, $f(x) \propto x \sqrt{1-x^2} \, \exp[-\xi(1-x^2)]$.
       See H. Albrect {\it et al.} [The ARGUS Collaboration], Phys. Lett. {\bf B241},
       278 (1990).
    %
    \bibitem{pdg} K. Nakamura {\it et al.} [The Particle Data Group],
       J. Phys. {\bf G37}, 075021 (2010).
    %
    \bibitem{markIII} Z. Bai {\it et al.} [The Mark III Collaboration],
       Phys. Rev. Lett, {\bf 65}, 1309 (1990).
    %
    \bibitem{BES} M. Ablikim {\it et al.} [The BES Collaboration],
       Phys. Lett. {\bf B662}, 330 (2008).
    %
    \bibitem{splots} M. Pivk and F. R. Le Diberder, Nucl. Instrum. Meth. {\bf A555}, 356 (2005).
    %
   \end{thebibliography}
\end{document}